\documentclass[preprint,showpacs,preprintnumbers,amsmath,amssymb,natbib]{revtex4}

\usepackage{graphicx}
\usepackage{epsfig}
\usepackage{dcolumn}
\usepackage{bm}

\def\be{\begin{equation}}
\def\ee{\end{equation}}
\def\ba{\begin{eqnarray}}
\def\ea{\end{eqnarray}}

\begin{document}

\title{Bell operator and Gaussian squeezed states in noncommutative quantum mechanics}

\author{Catarina Bastos\footnote{E-mail: catarina.bastos@ist.utl.pt}}
\affiliation{GoLP/Instituto de Plasmas e Fus\~ao Nuclear, Instituto Superior T\'ecnico, Universidade de Lisboa, Avenida Rovisco Pais 1, 1049-001 Lisboa, Portugal}

\author{Alex E. Bernardini\footnote{ E-mail: alexeb@ufscar.br}}
\affiliation{Departamento de F\'isica, Universidade Federal de S\~ao Carlos, PO Box 676, 13565-905, S\~ao Carlos, SP, Brasil.}

\author{Orfeu Bertolami\footnote{Also at Centro de F\'isica do Porto, Rua do Campo Alegre, 687, 4169-007 Porto, Portugal. E-mail: orfeu.bertolami@fc.up.pt}}
\affiliation{Departamento de F\'isica e Astronomia, Faculdade de Ci\^encias da Universidade do Porto, Rua do Campo Alegre, 687, 4169-007 Porto, Portugal}

\author{{Nuno Costa Dias}\footnote{Also at Grupo de F\'{\i}sica Matem\'atica, UL,
Avenida Prof. Gama Pinto 2, 1649-003, Lisboa, Portugal. E-mail: ncdias@meo.pt}}
\affiliation{Departamento de Matem\'{a}tica, Universidade Lus\'ofona de
Humanidades e Tecnologias Avenida Campo Grande, 376, 1749-024 Lisboa, Portugal}

\author{{Jo\~ao Nuno Prata}\footnote{E-mail: joaoprata@enautica.pt}}
\affiliation{Escola Superior N\'autica Infante D. Henrique, Av. Eng. Bonneville Franco\\
2770-058 Pa\c{c}o de Arcos\\
Portugal}
\date{\today}

\begin{abstract}
One examines putative corrections to the Bell operator due to the noncommutativity in the phase-space. 
Starting from a Gaussian squeezed envelop whose time evolution is driven by commutative (standard quantum mechanics) and noncommutative dynamics respectively, one concludes that, although the time evolving covariance matrix in the noncommutative case is different from the standard case, the squeezing parameter dominates and there are no noticeable noncommutative corrections to the Bell operator. This indicates that, at least for squeezed states, the privileged states to test Bell correlations, noncommutativity versions of quantum mechnics remains as non-local as  quantum mechanics itself. 
\end{abstract}

\maketitle

\section{Introduction}

In their well known ``EPR" paper, Einstein, Podolsky and Rosen \cite{EPR} discussed the dynamical properties of a composite quantum system with two interacting subsystems.
Despite the originality of their approach, they controversially concluded that quantum mechanics was not complete and that some underlying (hidden) variables would be necessary for a complete physical description of reality.
They further argued that as, in order to predict the evolution of a physical quantity at a given precision, the system should not be disturbed.
However, it is known that if two physical quantities are described by noncommuting operators, then the knowledge of one observable precludes the knowledge of the other with an arbitrary precision. Thus, they concluded, either quantum mechanics does not describe reality in a complete fashion, or the two noncommuting quantities can be simultaneously described with an arbitrary precision.

In the 1950s, after the work of Bohm and Aharonov \cite{BA1}, spin-like systems have been studied, and considered as particularly suitable for testing the assumptions underlying a hidden variable hypothesis.
However, a crucial new element was introduced by John Bell \cite{Bell1}, who showed that quantum correlations are essentially non-local.
This allowed for a theoretical analysis of the hidden variable scenario, leading eventually to Bell's inequalities \cite{Bell1}.
However, the original formulation of the EPR discussion was based on continuous variable systems. Actually, quantum correlations for position-momentum variables can be studied in the phase-space using the the Wigner function formulation of quantum mechanics. Bell argued that, due to the positivity of the Wigner function, the original EPR states would be necessarily non-local \cite{Bell1}.
Recently, it has been shown that Wigner functions of two-mode squeezed vacuum states, being positive definite, provide a direct evidence of the non-locality of states \cite{WF}. The Wigner function can be interpreted as a correlation function for the joint measurement of the parity operator. Clearly, the connection of the Wigner function to the parity operator, in this case, is linked to the fact that a dichotomic operator is needed in order to study the quantum systems. The key point is that a state does not have to violate all possible Bell's inequalities to be non-local, but rather to violate just one of the Bell's inequalities. 

In addition, the definition of entanglement is that a composite system cannot be written as a product of states of individual subsystems. Thus, there exists a relation between entanglement and EPR correlations. However, there are Gaussian states that are entangled, but are compatible with a local hidden variable theory, and there are entangled Gaussian states that are non-local \cite{Buono1, Buono2}. This confirms the existence of different types of quantum correlations and non-locality.
Indeed, the role of quantum entanglement has been acknowledged for its wide range of applications in quantum information protocols \cite{theory2} and quantum communication \cite{theory1}. One of the key results is the so-called positive partial transposed (PPT) separability criterion \cite{Peres}, which provides a necessary and, in some cases, a sufficient condition for distinguishing between separable and entangled states in discrete quantum systems. This criterion was extended to continuous variable systems by implementing the partial transpose operation as a mirror reflection of the Wigner formulation in phase-space \cite{Simon}. The ``continuous'' PPT criterion has important applications in the theory of quantum information of Gaussian states \cite{gauss01,Adesso}, which is at the core of testing procedures for estimating quantum correlations \cite{Weedbrook}. In these cases, the PPT criterion yields both a necessary and sufficient separability condition
\cite{Adesso,Simon}. Gaussian states are also quite useful for investigating the entanglement exclusively induced by a noncommutative (NC) deformation of phase-space \cite{Bastos6,Alex01,Alex02}.

Before discussing these issues in the context of phase-space noncommutative quantum mechanics (NCQM), let us point out that configuration space noncommutatitvity is believed to be an intrinsic feature of quantum gravity and string theory \cite{Connes,Seiberg}. Phase-space noncommutativity, on its turn,  has shown to have striking features, with implications for quantum cosmology, black hole singularity and thermodynamics \cite{Bastos3}, and the equivalence principle \cite{Bastos4}. Furthermore, the quantum mechanical aspects of NC theories have focused on studies of the quantum Hall effect \cite{Prange,Belissard}, the gravitational quantum well for ultra-cold neutrons \cite{06A}, the Landau/$2D$-oscillator problem in the phase-space \cite{Nekrasov01,Rosenbaum}, for the graphene \cite{Bastos5}, and as a probe of quantum beating and missing information effects \cite{2013}.
Furthermore, the phase-space noncommutativity was shown to give rise to possible violations of the Robertson-Schr\"odinger uncertainty principle \cite{Bastos1}. It has also effects for the Ozawa's measurement-disturbance uncertainty principle, as phase-space noncommutativity introduces extra terms and turns, for instance, a backaction evading measurement into a non backaction evading one \cite{Bastos7}.

Phase-space NCQM are based on extensions of the Heisenberg-Weyl algebra \cite{GossonAdhikari,08A} and can be formulated in terms of phase-space Wigner functions \cite{Bastos0}. Here, a different approach is considered to introduce the noncommutativity. Instead of noncommutativity been considered as a feature of phase-space, it is a consequence of the interaction between the two modes, ``Alice" and ``Bob". In Ref. \cite{Bastos6} it was shown that the entanglement arises from the interaction between the two modes and that this entanglement has inherent NC feaures. Thus, it is both theoretically and experimentally interesting to test if an interaction between modes can be related to some form of noncommutativity between them. To study these correlations between modes, one considers NC version of the Bell operator defined, for continuous variable systems, in terms of the Wigner function of the system \cite{WF}.

In this framework the particular cases of Gaussian squeezed envelopes for time evolving Gaussian states \cite{Bastos6} deformed by the NCQM are considered.
On one hand, one considers a departure Gaussian squeezed envelop with time evolution driven either by commutative (standard quantum mechanics) or by NC dynamics.
Although the covariance matrix of the NC two mode time evolving Gaussian squeezed state is different from the commutative one, it is found that the squeezing parameter is the unique driving vector of quantum correlations and non-locality and it is not influenced by noncommutativity.


\section{Noncommutative Bell operator }

Let one considers a bipartite quantum system ($2-$dimensional system) described in terms of a subsystem A (Alice) and a subsystem B (Bob).
One may write the collective degrees of freedom of the composite system $\widehat{z} = (\widehat{z}^A,\,\widehat{z}^B)$, where $\widehat{z}^A =(\widehat{x}, \widehat{p}_x)$ and $\widehat{z}^B=(\widehat{y}, \widehat{p}_y)$ as the corresponding generalized variables of the two subsystems, which obey the commutation relations \cite{Bastos6}
\begin{equation}
\left[\widehat{z}^A_i, \widehat{z}^B_j \right] = i \Omega_{ij}, \hspace{1cm} i,j = 1, 2,
\label{eq1}
\end{equation}
where the associated matrix is given by ${\bf \Omega} = \left[\Omega_{ij} \right]$, being a real skew-symmetric non-singular $4 \times 4$ matrix of the form
\begin{equation}
{\bf \Omega} = \left(
\begin{array}{c c}
{\bf J} & {\bf \Upsilon}\\
- {\bf \Upsilon} & {\bf J}
\end{array}
\right),
\label{eq2}
\end{equation}
where ${\bf\Upsilon}$ measures the noncommutativity of the position and momentum,
\begin{equation}
{\bf \Upsilon} = \left(
\begin{array}{c c}
{\theta} & {0}\\
{0} & {\eta}
\end{array}
\right),
\label{eq2a}
\end{equation}
${\bf J}$ is the standard symplectic matrix
\begin{equation}
{\bf J} = \left(
\begin{array}{c c}
{0} & {1}\\
- {1} & {0}
\end{array}
\right),
\label{eq2b}
\end{equation}
and one uses natural units, where $\hbar =1 $.

The NC structure can be formulated in terms of commuting variables by considering a linear Darboux transformation (DT), $\widehat{z} = {\bf D} \widehat{\xi}$, where $ \widehat{\xi}=( \widehat{\xi}^A,
\widehat{\xi}^B)$, with $ \widehat{\xi}^A= ( \widehat{x}_c, \widehat{p_x}_c)$ and $ \widehat{\xi}^B= ( \widehat{y}_c,\widehat{p_y}_c)$, satisfy the usual commutation relations \cite{Bastos0}:
\begin{equation}
\left[\widehat{\xi}_i, \widehat{\xi}_j \right] = i J_{ij} , \hspace{1cm} i,j=1, 2.
\label{eq3}
\end{equation}
where $[J_{ij}]={\bf J}$ from Eq. (\ref{eq2b}). The linear transformation ${\bf D} \in Gl(4)$ is such
that ${\bf\Omega} ={\bf D} {\bf J} {\bf D}^T  $. Notice that the map ${\bf D}$ is not uniquely defined. If one composes ${\bf D}$ with block-diagonal canonical transformations one obtains an equally valid DT. The matrix form of the DT is, therefore,
\be\label{DT}
{\bf D}=\left(
\begin{array}{c c c c}
\lambda &  0 &  0 & -{\theta\over 2\lambda}\\
0 &  \mu &  {\eta\over 2\mu} & 0\\
0 &  {\theta\over 2\lambda} & \lambda & 0\\
-{\eta\over 2\mu} & 0 &  0 & \mu
\end{array}\right)~,
\ee
where $\lambda$ and $\mu$ are arbitrary real parameters.

In order to obtain the non-locality of a state through the Clauser, Horne, Shimony and Holt (CHSH) inequality \cite{CHSH}, one must, in general, have an explicit definition for the Bell operator. This operator usually represents a combination of dichotomy measurements. The result is that, if the expectation value of such a Bell operator violates the corresponding inequality, then the system is non-local. In continuous variable quantum information, in order to mimic the dichotomic behavior, the parity operator is used. The later can be determined, on the photon number, by assigning $+1$ or $-1$, depending on whether an even or an odd number of photons are registered. For the connection between  the Wigner function of the state and the joint measurement of the parity operator on the bipartite quantum state, see e.g. Ref. \cite{WF}.

Following Ref. \cite{WF}, the Bell operator in the commutative case is given by the linear combination of four expectation values,
\be\label{Bell1}
\mathcal{B}=\langle \mathcal{M}(0,0)\rangle + \langle \mathcal{M}(\alpha_1,0)\rangle + \langle \mathcal{M}(0,\alpha_2)\rangle -\langle \mathcal{M}(\alpha_1,\alpha_2)\rangle~,
\ee
where $\alpha_{1,2}$ are the phase-space variables (or symbols) associated with the operators $\hat{\alpha}_{1}=\hat{x}_{c }+i \hat{p_x}_{c}$ and $\hat{\alpha}_{2}=\hat{y}_{c}+i \hat{p_y}_{c}$ which are complex amplitudes that carry two degrees of freedom, $\hat{\alpha}$ and  $\hat{\alpha}^*$ for each mode, $1$ and $2$, and
\be\label{Bell2}
\langle \mathcal{M}(\alpha_1,\alpha_2)\rangle\equiv {\pi^2\over 4} W(\alpha_1,\alpha_2) ,
\ee
with $W(\alpha_1,\alpha_2)$ being the commutative Wigner function of the state calculated in $(\alpha_1,\alpha_2)$. 
Local theories admit a description in terms of local hidden variables \cite{WF}, i.e. $\alpha_{1}$ and $\alpha_{2}$ (in Eq.(\ref{Bell1})).This variables are the ones that will be used to paramterize the dichotomic behaviour  of the Bell operator. Let us consider a free parameter $\mathcal{I}$ that will represent this dichotomic behaviour. In that sense, $\alpha_{1}$ and $\alpha_{2}$, can only correspond to $\mathcal{I}$ or $-\mathcal{I}$.
On the other hand, non-locality is identified by 
\be\label{Bell3}
|\mathcal{B}_{min}| > 2~.
\ee

From Eqs. (\ref{Bell1}) and (\ref{Bell2}), one notices that the Bell operator is a functional depending on the Wigner function of the state. In order to understand what is the real effect of NC in the Bell functional, let one considers an analogy with what happens to the expectation values of the position, $\langle x \rangle$, which are also functionals of the Wigner function. 
One begins the analysis with a Wigner function that could be either commutative (i. e. standard QM), $W(x,p)$, or NC, $W^{NC}(x,p)$ \cite{Bastos0}. That is, due to the fact that one actually does not know if the state obeys a NC dynamics or not. The only requirement here is that the initial Wigner function, $W_0(x,p)$, will give the expectation values that are going to be observed in the laboratory at the initial time. The expectation value $<x>$ is also given by a functional of the Wigner function, that is $\langle x\rangle=\int{dx~ dp~ x~ F(x,p)}$, which is the same in the commutative and in the NC case. Thus, it is expected that at the initial time, $W^C_0(x,p)=W^{NC}_0(x,p)$. Analogously, one should have $\mathcal {B}^{NC}_0\equiv \mathcal{B}(W^{NC}_0(x,p))=\mathcal{B}(W^{C}_0(x,p))\equiv \mathcal{B}^C_0$.

Now, to understand what is the influence of NC in the expectation value, $\langle x \rangle$, the evolution of the system with the two Hamiltonians, the commutative one and the NC one, should be studied. Then, one should look for discrepancies between the values of the expectation value $\langle x_t\rangle$ in the two cases. This clearly can be done in the laboratory in order to determine if the state is either NC or commutative, once one lets the system evolve in time. In this case, $\mathcal{B}^{NC}(t)=\mathcal{B}(W^{NC}_t(x,p))$ and one should compare it with $\mathcal{B}(t)=\mathcal{B}(W_t(x,p))$. 

In what follows, we considere a NC Gaussian squeezed state and compute the Bell functionals for the NC and commutative case for the two states. For Gaussian states, in cases in which the CHSH inequality can be rewritten in terms of Gaussian symplectic invariants \cite{CHSH,Ball70}, the noncommutativity may change the Gaussian covariant matrix without changing their invariants. 

Of course, it might exist other ways to implement NC effects into the Bell operator, however we shall not venture here into this path and will constrain ourselves to examine just squeezed states.

\section{NC two mode squeezed state}

One considers now a bipartite Gaussian state \cite{Buono1},
\be
\label{Wigner1}
W({\bf{R}})=\frac{1}{\pi^2{\sqrt{Det[\sigma]}}}{\exp{\left(-{1\over2}{\bf{R}}^{T}\sigma^{-1}{\bf{R}}\right)}}~,
\ee
where ${\bf R}\equiv(x,p_x,y,p_y)$ is the vector of a set of orthogonal quadratures, for modes {\it a} and {\it b}, respectively. The covariance matrix, $\sigma$, is given by a $4\times4$ matrix written in the following form,
\be\label{CM1}
{\sigma}=\left(
\begin{array}{c c }
\bf\alpha &  \bf\gamma \\
\bf\gamma^{T} &  \bf\beta
\end{array}\right),
\ee
where $\bf\alpha$ and $\bf\beta$ represent the self-correlation of each single subsystem and $\bf\gamma$ describes the correlation between the two subsystems. Then, any covariance matrix that describes a physical state can be written, through a local symplectic transformation, in the standard form:
\be\label{CM2}
{\sigma}=\left(
\begin{array}{c c c c}
n &  0 &  c_1 & 0\\
0 &  n &  0 & c_2\\
c_1 &  0 & m & 0\\
0 & c_2 &  0 & m
\end{array}\right)~,
\ee
where $n, m, c_1$ and $c_2$ are determined by the four local symplectic invariants $I_1\equiv {Det[\bf\alpha]}=n^2$, $I_2\equiv {Det[\bf\beta]}=m^2$, $I_3\equiv {Det[\bf\gamma]}=c_1c_2$ and $I_4\equiv {Det[\sigma]}=(nm-c_1^2)(nm-c_2^2)$.

When considering Bell inequalities, most often a squeezed state is used.
By considering the case of a squeezed state in standard commutative quantum mechanics, a departure Hamiltonian can be given by \cite{Braunstein}
\be\label{SqueezHam}
H=ik (\zeta^* a^{\dagger}b^{\dagger}-  \zeta a b)~,
\ee
where $a^{\dagger}, b^{\dagger}, a, b$ are the creation and annihilation operators of the two modes, A and B, respectively. The only non-vanishing commutators are $[a, a^{\dagger}]=[b, b^{\dagger}]=1$. Moreover, $\zeta=r e^{-i \phi}$ , where $r$ is the squeezing parameter and $\phi$ a phase associated to the ``pump" that produces the squeezed state. If the phase is null, $\phi=0$, then the equations of motion are solved for the operators, $\dot{O}=-i [O,H]$, and one obtains the way the operators transform \cite{Braunstein}:
\ba\label{SqueezTransf}
a= a_{0} \cosh(r) +b_{0}^{\dagger} \sinh(r)\hspace{0.2cm}&,&\hspace{0.2cm} a^{\dagger}= a_{0}^{\dagger} \cosh(r)+b_{0} \sinh(r)~,\nonumber\\
b= b_{0} \cosh(r) +a_{0}^{\dagger} \sinh(r)\hspace{0.2cm}&,&\hspace{0.2cm} b^{\dagger}= b_{0}^{\dagger} \cosh(r)+a_{0} \sinh(r)~.
\ea
This is the how creation/annihilation operators transform when one applies a squeezing transformation, given by the operator $S=\exp{(\zeta^* ab-\zeta a^{\dagger}b^{\dagger})}$, where the quadratures are defined as
\ba\label{quadratures1}
x={1\over\sqrt{2}} (a+a^{\dagger})\hspace{0.2cm}&,&\hspace{0.2cm}p_x={-i\over \sqrt{2}}(a-a^{\dagger})~,\nonumber\\
y={1\over\sqrt{2}} (b+b^{\dagger})\hspace{0.2cm}&,&\hspace{0.2cm}p_y={-i\over \sqrt{2}}(b-b^{\dagger})~.
\ea
Applying the Bogoliubov transformation,
\ba\label{quadratures2}
x= x_0\cosh(r)+y_0\sinh(r)\hspace{0.2cm}&,&\hspace{0.2cm} p_x=p_{x_0}\cosh(r)-p_{y_0}\sinh(r)~,\nonumber\\
y= y_0\cosh(r)+x_0\sinh(r)\hspace{0.2cm}&,&\hspace{0.2cm} p_y=p_{y_0}\cosh(r)-p_{x_0}\sinh(r)~,
\ea
one is able to define ${\bf S}$ as the matrix that transforms the quadratures into the quadratures of a squeezing state:
\be\label{Squeez2}
{\bf S}(r)=\left(
\begin{array}{c c c c}
\cosh(r) &  0 &  \sinh(r) & 0\\
0 &  \cosh(r) &  0 & -\sinh(r)\\
\sinh(r) &  0 & \cosh(r) & 0\\
0 & -\sinh(r) &  0 & \cosh(r)
\end{array}\right)~,
\ee
where the time evolution is given in terms of the squeezing parameter $r$, and which enables one to construct the covariance matrix of the squeezed state, ${\bf \Sigma}(r)={\bf S^T}(r)\,{\bf S}(r)$. 

The same strategy can be applied for the NC case, but now one has to take into account that the position and momenta obey the NC algebra: 
\begin{equation}
\left[x, y \right] = i \theta\hspace{0.2cm},\hspace{0.2cm}\left[p_x, p_y \right] = i \eta\hspace{0.2cm},\hspace{0.2cm}\left[x, p_x \right] =\left[y, p_y\right]= i~,
\label{NCQM}
\end{equation}
where $\theta$ and $\eta$ are new fundamental constants of Nature. The NC variables can be related with the commutative ones by a Darboux transformation (or Seiberg-Witten map), Eq. (\ref{DT}), $\hat{z}={\bf D}\hat{\xi}$.
In particular, in this case the NC variables are given by,
\ba\label{Dmap}
x=\lambda x_c-{\theta\over 2\lambda}p_{y_c}\hspace{0.2cm},\hspace{0.2cm} p_{x}=\mu p_{x_c}+{\eta\over 2\mu} y_c~,\nonumber\\
y=\lambda y_c+{\theta\over 2\lambda}p_{x_c}\hspace{0.2cm},\hspace{0.2cm} p_{y}=\mu p_{y_c}-{\eta\over 2\mu} x_c~,
\ea
where $\lambda$ and $\mu$ are dimensionless constants, such that $2\lambda\mu= 1+\sqrt{1-\theta\eta}$ so to ensure that the map is invertible, and the index $c$ denotes commutative variables.
Through the definition of creation/annihilation operators in terms of position and momenta $a=(1/\sqrt{2})(x+ip_y)$, $a^{\dagger}=(1/\sqrt{2})(x-ip_y)$ (analogous to the B mode), one obtains
\ba\label{operatorsNC}
a={1\over\sqrt{2}}\left[\lambda x_c-{\theta\over 2\lambda}p_{y_c}+i \left(\mu p_{x_c}+{\eta\over 2\mu}y_c\right)\right]~,\\
a^{\dagger}={1\over\sqrt{2}}\left[\lambda x_c-{\theta\over 2\lambda}p_{y_c}-i \left(\mu p_{x_c}+{\eta\over 2\mu}y_c\right)\right]~,\\
b={1\over\sqrt{2}}\left[\lambda y_c+{\theta\over 2\lambda}p_{x_c}+i \left(\mu p_{y_c}-{\eta\over 2\mu}x_c\right)\right]~,\\
b^{\dagger}={1\over\sqrt{2}}\left[\lambda y_c+{\theta\over 2\lambda}p_{x_c}-i \left(\mu p_{y_c}-{\eta\over 2\mu}x_c\right)\right]~.
\ea
Then, one finds the NC quadrature operators solving the previous system of equations,
\ba\label{quadraturesNC}
x_c={\mu\over {\sqrt{2}(2\lambda\mu-1)\mu}} \left[(a+a^{\dagger})-i {\theta\over 2\lambda\mu}(b-b^{\dagger})\right]~,\\
p_{x_c}={1\over {2\sqrt{2}(2\lambda\mu-1)\mu}} \left[i (a^{\dagger}-a)-\eta(2\lambda\mu)(b+b^{\dagger})\right]~,\\
y_c={\mu\over {\sqrt{2}(2\lambda\mu-1)\mu}} \left[(b+b^{\dagger})+i {\theta\over 2\lambda\mu}(a-a^{\dagger})\right]~,\\
p_{y_c}={1\over {2\sqrt{2}(2\lambda\mu-1)\mu}} \left[i (b^{\dagger}-b)-\eta(2\lambda\mu)(a+a^{\dagger})\right]~.
\ea 
Since now the positions and the momenta do not commute in the phase space, one evaluates all commutators for the creation and annihilation operators in terms of the commutative variables
$[a, a^{\dagger}]=[b, b^{\dagger}]=1$, $[a, b] =[a^{\dagger}, b^{\dagger}]={i \over 2}(\theta-\eta)$, $[a, b^{\dagger}]=[a^{\dagger}, b]={i \over 2}(\theta+\eta)$, and the rest of the commutators vanish. The equations of motion for the creation and annihilation operators become
\ba\label{eqsmotionNC}
{da\over dt}={1\over i} [a,H]=k \left[ b^{\dagger}+{i\over2}\left(\theta (a^{\dagger}-a)+\eta(a^{\dagger}+a)\right)\right]~,\\
{da^{\dagger}\over dt}={1\over i} [a^{\dagger},H]=k \left[ b+{i\over2}\left(\theta (a^{\dagger}-a)-\eta(a^{\dagger}+a)\right)\right]~,\\
{db\over dt}={1\over i} [b,H]=k \left[ a^{\dagger}-{i\over2}\left(\theta (b^{\dagger}-b)+\eta(b^{\dagger}+b)\right)\right]~,\\
{db^{\dagger}\over dt}={1\over i} [b^{\dagger},H]=k \left[ b^{\dagger}+{i\over2}\left(\theta (a^{\dagger}-a)+\eta(a^{\dagger}+a)\right)\right]~.
\ea
Solving the system of equations, considering $kt=r$, where $r$ is the squeezing parameter, and using the definitions of the quadratures operators, one has the following Bogoliubov transformation:
\ba\label{BogoliubovNC}
x=\cosh {(\sqrt{\theta\eta} r)}\left[x_0 \cosh(r)+y_0 \sinh(r)\right]+\sqrt{\theta\over\eta} \sinh{(\sqrt{\theta\eta}r)}\left[p_{x_0} \cosh(r)-p_{y_0} \sinh(r)\right]~,\\
p_x=\cosh {(\sqrt{\theta\eta} r)}\left[p_{x_0} \cosh(r)-p_{y_0} \sinh(r)\right]+\sqrt{\eta\over\theta} \sinh{(\sqrt{\theta\eta}r)}\left[{x_0} \cosh(r)+{y_0} \sinh(r)\right]~,\\
y=\cosh {(\sqrt{\theta\eta} r)}\left[y_0 \cosh(r)+x_0 \sinh(r)\right]-\sqrt{\theta\over\eta} \sinh{(\sqrt{\theta\eta}r)}\left[p_{y_0} \cosh(r)-p_{x_0} \sinh(r)\right]~,\\
p_y=\cosh {(\sqrt{\theta\eta} r)}\left[p_{y_0} \cosh(r)-p_{x_0} \sinh(r)\right]-\sqrt{\eta\over\theta} \sinh{(\sqrt{\theta\eta}r)}\left[{y_0} \cosh(r)+{x_0} \sinh(r)\right]~.
\ea
Thus, in order to have the NC covariance matrix one needs 
\footnotesize\be\label{Squeez1NC}
{\bf S}^{NC}(r)=\left(
\begin{array}{c c c c}
\cosh{(\xi r)}\cosh(r) &  \sqrt{\theta\over\eta}\sinh{(\xi r)}\cosh(r) & \cosh{(\xi r)} \sinh(r) & -\sqrt{\theta\over\eta}\sinh{(\xi r)}\sinh(r)\\
\sqrt{\eta\over\theta}\sinh{(\xi r)}\cosh(r) & \cosh{(\xi r)} \cosh(r) &  \sqrt{\eta\over\theta}\sinh{(\xi r)}\sinh(r) & -\cosh{(\xi r)}\sinh(r)\\
\cosh{(\xi r)}\sinh(r) &  \sqrt{\theta\over\eta}\sinh{(\xi r)}\sinh(r) &\cosh{(\xi r)} \cosh(r) &  -\sqrt{\theta\over\eta}\sinh{(\xi r)}\cosh(r)\\
-\sqrt{\eta\over\theta}\sinh{(\xi r)}\sinh(r) & -\cosh{(\xi r)}\sinh(r) &   -\sqrt{\eta\over\theta}\sinh{(\xi r)}\cosh(r) & \cosh{(\xi r)}\cosh(r)
\end{array}\right).
\ee\normalsize
Hence, likewise the commutative case, one can obtain the covariance matrix, ${\bf\Sigma}^{NC}(r)={\bf {S^{NC}}^T}(r){\bf S^{NC}}(r)$. 
Besides having the same departure Gaussian states (envelops), i. e. ${\bf\Sigma}^{NC}(0)={\bf\Sigma}(0)$ since ${\bf S^{NC}}(0) = {\bf S}(0)$, which guarantees that the NC effects are purely due to the dynamics driven by the corresponding (non)commutative Hamiltonian,
one notices that the symplectic invariants from Eq. (\ref{Squeez1NC}) are the same as those from Eq. (\ref{Squeez2}), such that since the covariance matrix,  ${\bf\Sigma}^{NC}(r)={\bf {S^{NC}}^T}(r){\bf S^{NC}}(r)$ is identified with $\sigma$ from Eq.~(\ref{Wigner1}), the second moments of the Gaussian state (c.f. Eq.~(\ref{Wigner1})) are not affected by the NC map.

\section {Bell operator of a NC squeezing state in NC Quantum Mechanics}

Knowing the covariance matrix for the NC squeezed state one is now able to evaluate the Wigner function for the quantum state. The inverse of the covariance matrix is written as
\be\label{Squeez2NC}
({\bf \Sigma}^{NC})^{-1}=\left(
\begin{array}{c c c c}
n & d & t_1 & -c \\
d & m & c & t_2 \\
t_1 & c & n & -d\\
-c & t_2 & -d & m
\end{array}\right).
\ee
where the parameter $r$ was suppressed from the notation, such that
\ba\label{Squeez3NC}
n&=& {\cosh(2r)\over 2} \left[(1+\cosh{(2\xi r)})-{\theta\over \eta} (1-\cosh{(2\xi r)})\right]~,\nonumber\\
m&=&{\cosh(2r)\over 2} \left[(1+\cosh{(2\xi r)})-{\eta\over \theta} (1-\cosh{(2\xi r)})\right]~,\nonumber\\
d&=&-{{(\theta+\eta)}\over 2\xi}\cosh(2r)\sinh{(2\xi r)}~,\nonumber\\
c&=& {{(\theta+\eta)}\over 2\xi}\sinh(2r)\sinh{(2\xi r)}~,\nonumber\\
t_1&=&-{\sinh(2r)\over 2} \left[(1+\cosh{(2\xi r)})-{\theta\over\eta}(1-\cosh{(2\xi r)})\right]\nonumber\\
t_2&=&{\sinh(2r)\over 2} \left[(1+\cosh{(2\xi r)})-{\eta\over\theta}(1-\cosh{(2\xi r)})\right]~.
\ea

After some algebraic manipulation, the NC Wigner function Eq. (\ref{Wigner1}) can be written as
\ba\label{NCWF1}
W(\alpha_1,\alpha_2)={1\over \pi^2}\exp\{&-&{1\over4}\left[(n-m)(|\alpha_1|^2+|\alpha_2|^2)+2c(\alpha_1\alpha_2^*-\alpha_1^*\alpha_2)+\right.\nonumber\\
&+&{1\over2}\left((n+m+2d)(\alpha_1^2+(\alpha_2^*)^2)+(n+m-2d)((\alpha_1^*)^2+\alpha_2^2)\right)+\nonumber\\
&+&\left.(t_1+t_2)(\alpha_1\alpha_2+\alpha_1^*\alpha_2^*)+(t_1-t_2)(\alpha_1\alpha_2^*+\alpha_1^*\alpha_2)\right]\}~,
\ea
where $\alpha_1=x_0+ip_{x_0}$ and $\alpha_2=y_0+ip_{y_0}$ are complex amplitudes. Starting from Eq. (\ref{Bell1}) and using the definitions, Eqs. (\ref{Bell2}) and (\ref{NCWF1}), one otains after some algebraic manipulation that
\be\label{BellNCsq1}
\mathcal{B}={1\over 4}\left[1+2 \exp{\left(-{n\over2}\mathcal{I}\right)}-\exp{\left(-(n-t_1)\mathcal{I}\right)}\right]~,
\ee
where $\mathcal{I}$ is some complex displacement amplitude. Thus, one has a Bell operator, $\mathcal{B}(\mathcal{I},n,t_1)$, that depends on a free parameter $\mathcal{I}$, and on the state properties $n$ and $t_1$. Looking for the maximum violation for a given state, one should maximize the Bell operator in terms of the free parameter $\mathcal{I}$ \cite{Buono1},
\be\label{maxBellNC}
{\partial\mathcal{B}\over \partial\mathcal{I}}|_{\mathcal{I=\tilde{\mathcal{I}}}}=0 \Leftrightarrow \tilde{\mathcal{I}}={2\over n-2t_1} \ln{\left({n-t_1\over n}\right)}~.
\ee
which leads to
\be\label{maxBellNC2}
\mathcal{B}(\tilde{\mathcal{I}}) = {1 \over 4}
\left[1+ \frac{(3n-2t_1)\left(1-\frac{t_1}{n}\right)^{-\frac{n}{n-2t_1}}}{n-t_1}\right].
\ee
Substituting $n$ and $t_1$ by the respective functions, Eqs.~(\ref{Squeez3NC}), one finally obtains 
\be
\tilde{\mathcal{I}}=\tilde{\mathcal{I}}(\theta,\eta) = \frac{4 \eta \log\left(1 + \tanh(2r)\right)}{\eta - \theta + (\eta + \theta) \cosh(2 \xi r) (\cosh(2r) + 2 \sinh(2r))}
\ee     
where the NC dynamics engenders the dependence of $\tilde{\mathcal{I}}$ on $\theta$ and $\eta$.

The result for the Bell operator is given by
\be
\mathcal{B}(\tilde{\mathcal{I}}(\theta,\eta)) = {1\over 4}
\left[1 + \frac{\exp(-2 r) (3 \cosh(2r) + 2 \sinh(2r))}{(1 + \tanh(2r))^{\frac{1}{1 + 2 \tanh(2r)}}}\right],
\ee
which implies that, it is exactly identical to the Bell operator in the standard QM (commutative) limit, $  \tilde{\mathcal{I}}(\theta,\eta) \to \tilde{\mathcal{I}}(0,0)$, where:  
\be
\tilde{\mathcal{I}}(0,0) = \frac{2\log\left(1 + \tanh(2r)\right)}{\cosh(2r) + 2 \sinh(2r)},
\ee     

In order to admit local theories in terms of hidden variables this Bell operator should satisfy the condition
\be\label{BellNCsq3}
|\mathcal{B}(\tilde{\mathcal{I}})|\leq 2~.
\ee

Thus, for a set of departure Gaussian states for which $\mathcal {B}^{NC}_0\equiv \mathcal{B}(W^{NC}_0(x,p))=\mathcal{B}(W^{C}_0(x,p))\equiv \mathcal{B}^C_0$, one concludes that the NC Bell operator for the squeezed state that depends on the NC parameters, $\mathcal{B}(\tilde{\mathcal{I}}(\theta,\eta)$, is the same as the Bell operator in the commutative case, $\mathcal{B}(\tilde{\mathcal{I}}(0,0)$. Here, the squeezing parameter dominates and the noncommutativity effectively does not interfere on the squeezing state engendered by Eq.~(\ref{Wigner1}) once it is driven by the Hamiltonian, Eq.~(\ref{SqueezHam}).


\section{Conclusions}

In this work it is shown that for squeezed states phase noncommutativity does not introduce corrections to the Bell operator. In order to show that one considers, in particular, a noncommutativity between modes as this is a way to strengthen the interaction between them and has been considered earlier when considering an entanglement state \cite{Bastos6}. However, when considering a NC squeezed state, it is shown that, despite the NC corrections in the evolution of the startes, the squeezing parameter is dominant and the Bell operator does not change when the noncommutativity is introduced.  This result seems to indicate that the non-localities that the Bell operator reveals for quantum mechanics are unaltered by noncommutativity. One way to interpret  this result is that noncommutativity affects the way a system evolves, through the changes in the Hamiltonian, although not necessarily introducing further non-localities to the quantum mechanical problem. Of course, one could conjecture that this result might be changed by going beyond squeezed states or by considering more general (non canonical) NC structures.. For sure, this would open interesting experimental opportunities to test noncommutativity and certainly deserve some further study.

\begin{acknowledgements}
This work is supported by the COST action MP1405. The work of CB is supported by the FCT (Portugal) grant SFRH/BPD/62861/2009. The work of AEB is supported by the Brazilian Agencies FAPESP (grant 15/05903-4) and CNPq (grant 300809/2013-1 and grant 440446/2014-7).
 \end{acknowledgements}

\end{document}